\documentclass[12pt]{article}
\usepackage{amsmath,epsf,amssymb,latexsym,cite,xy}
\xyoption{matrix}\xyoption{arrow}
\newtheorem{conjecture}{Conjecture}

\newif\iffigs\figstrue

\DeclareFontFamily{U}{rsf}{}
\DeclareFontShape{U}{rsf}{m}{n}{
  <5> <6> rsfs5 <7> <8> <9> rsfs7 <10-> rsfs10}{}
\DeclareMathAlphabet\Scr{U}{rsf}{m}{n}

\def\pplogo{\vbox{\kern-\headheight\kern -29pt
\halign{##&##\hfil\cr&{%\sc
\ppnumber}\cr\rule{0pt}{2.5ex}&\ppdate\cr}
}}
\makeatletter
\def\ps@firstpage{\ps@empty \def\@oddhead{\hss\pplogo}%
  \let\@evenhead\@oddhead % in case an article starts on a left-hand page
}
\def\maketitle{\par
 \begingroup
 \def\thefootnote{\fnsymbol{footnote}}
 \def\@makefnmark{\hbox{$^{\@thefnmark}$\hss}}
 \if@twocolumn
 \twocolumn[\@maketitle]
 \else \newpage
 \global\@topnum\z@ \@maketitle \fi\thispagestyle{firstpage}\@thanks
 \endgroup
 \setcounter{footnote}{0}
 \let\maketitle\relax
 \let\@maketitle\relax
 \gdef\@thanks{}\gdef\@author{}\gdef\@title{}\let\thanks\relax}
\makeatother

\def\O{{\mathcal O}}
\def\C{{\mathbb C}}
\def\P{{\mathbb P}}

\def\R{{\mathbb R}}
\def\Z{{\mathbb Z}}

\def\Hom{\operatorname{Hom}}

\def\Spec{\operatorname{Spec}}

\def\GU{\operatorname{U{}}}

\def\ch{\operatorname{\mathit{ch}}}
\def\td{\operatorname{\mathit{td}}}

\def\CY{Calabi--Yau}

\def\cM{{\Scr M}}

\def\cK{{\Scr K}}

\def\cT{{\Scr T}}

\def\cE{{\Scr E}}
\def\cF{{\Scr F}}

\def\DC{\mathbf{D}}
\def\ff#1#2{{\textstyle\frac{#1}{#2}}}

\def\labto#1{\mathrel{\mathop\to^{#1}}}
\def\Lotimes{\mathrel{\mathop\otimes^{\mathbf{L}}}}

\begin{document}
\setcounter{page}0
\def\ppnumber{\vbox{\baselineskip14pt
\hbox{DUKE-CGTP-01-01}
\hbox{NSF-ITP-01-11}
\hbox{hep-th/0102198}}}
\def\ppdate{February 2001} \date{}

\title{\LARGE Some Navigation Rules for D-Brane Monodromy\\[10mm]}
\author{
Paul S. Aspinwall\\[10mm]
\normalsize Center for Geometry and Theoretical Physics, \\
\normalsize Box 90318, \\
\normalsize Duke University, \\
\normalsize Durham, NC 27708-0318\\[10mm]
}

{\hfuzz=10cm\maketitle}

\def\Large{\large}
\def\LARGE{\large\bf}

%\vskip 1cm

\begin{abstract}
We explore some aspects of monodromies of D-branes in the K\"ahler moduli space
of \CY\ compactifications. Here a D-brane is viewed as an object of
the derived category of coherent sheaves. We compute all the
interesting monodromies in some nontrivial examples and
link our work to recent
results and conjectures concerning helices and mutations. We note some
particular properties of the 0-brane.
\end{abstract}

\vfil\break

%%%%%%%%%%%%%%%%%%%%%%%%%%%%%%%%%%%%%%%%%%%%%%%%%%%%%%%%%%%%%%%%

\section{Introduction}    \label{s:intro}

There has been something of an evolution in our ideas about how a
D-brane should be considered. For the purposes of this paper we are interested
only in the even-dimensional branes in a type II string (so-called
``B-branes''). 
The sequence of a ideas have progressed roughly as
follows
\begin{enumerate}
\item A D-brane is something on which an open string may end.
\item A D-brane is a $\GU(N)$ gauge theory living on a subspace with
scalar fields spanning the normal bundle.
\item A D-brane should be viewed as coming from K-theory \cite{MM:K,W:K}.
\item A D-brane should be viewed as an object of the derived
category\footnote{In this paper the derived category will always be
bounded at both ends.} of coherent sheaves \cite{Kon:mir,Doug:DC} (see
also \cite{Shrp:DC}).
\end{enumerate}
We could also add that for non trivial $H$ field a D-brane should be
viewed as an object of the derived 
category of sheaves of modules over an Azumaya algebra \cite{KO:Azum}.
We will assume $H$ is trivial and so view
D-Branes as an object of the derived category of coherent sheaves. We
will consider our target space to be a \CY\ threefold $X$ and we
denote the derived category in question as $\DC(X)$. For the purposes
of this paper we ignore any issues concerning the stability of
D-branes. Our D-branes, which are objects of $\DC(X)$, were called
``topological D-branes'' in \cite{Doug:DC} where issues of stability
were discussed.

The reason that $\DC(X)$ is ``better'' than K-theory is that it
contains so much more information. For example {\em any\/} 0-brane on
$X$ corresponds to the same single element of K-theory whereas the object of
$\DC(X)$ corresponding to a 0-brane knows {\em where\/} this point is. That
is, K-theory measures the charge of the D-brane but $\DC(X)$ tells us
more and possibly all we could wish to know about a particular
D-brane.

As well as being knowledgeable about D-branes, $\DC(X)$ is also very
knowledgeable about $X$ itself. This shouldn't be too surprising as if
we know about all the 0-branes on $X$ then we know about all the points on
$X$ and so we should know about $X$ itself. Indeed for a very large
class of algebraic varieties it was shown by Bondal and Orlov
\cite{BO:DCeq} that $X$ is completely determined, as an algebraic
variety, by $\DC(X)$. While this process doesn't quite work for \CY\
varieties it is easy to speculate that adding such data as
the spectrum of central charges while in a ``\CY\ phase'' may
provide the missing information. This would allow the target space to
be constructed only given worldsheet information. Clearly therefore
the derived category should be of great {\em physical\/} as well as
mathematical interest.

An interesting question, also studied in \cite{BO:DCeq}, concerns
in how many ways one may associate a derived category to a fixed
$X$. An ``autoequivalence'' of a derived category is a map from the
category to itself preserving all the intrinsic algebraic structure
associated to the category. Such a map need not preserve D-branes
themselves. For example an object representing a 2-brane may become
something which more resembles a 4-brane under such a transformation. In
terms of string theory, these autoequivalences can arise from monodromy in
the moduli space of the complexified K\"ahler form as
first observed by Kontsevich \cite{Kon:mir}. Indeed, the fact that
this monodromy action on $\DC(X)$ can be understood at all is one
the appealing aspects of the derived category. It was suggested in
\cite{AD:tang} that the derived category should play a r\^ole in the
heterotic string for similar reasons.

Since many interesting questions about $\DC(X)$ are associated to
these monodromies, the purpose of this paper is to explore some of the
aspects of these monodromies. The analysis of monodromies when the
moduli space of complexified K\"ahler forms has only one complex
dimension is pretty easy as we review in section \ref{s:one}. 
Most of the interesting properties of monodromy do not appear until
we explore higher-dimensional moduli spaces. We do this in the later
sections. 

One should note that many of these problems can be, and have
been, performed using the method of solving the Picard--Fuchs
equations and using analytic continuation (see
\cite{CDGP:,Mor:gid,CDFKM:I,CFKM:II,HKTY:} etc.). 
These methods have been used in the context of D-branes in such papers
as \cite{BDLR:Dq,DG:fracM,DDG:wrap,DR:Dell}.
Instead we will use
the language of derived categories where, we believe, the structure is
much simpler to understand. In this way, computation of the monodromy
is extremely easy (at least on the cohomology classes) and does not
require the aid of a computer.  
Note that we are not being at all
original in using the derived category --- computations along these
lines have been done in \cite{Kont:mon,ST:braid,Horj:DX,Hos:DX} for
example. Our work 
differs from the latter only in the way we probe more deeply into the
moduli space addressing such questions as monodromy around
Landau--Ginzburg points in multi-parameter examples. Monodromy using 
the derived category approach has also been studied recently in papers
such as \cite{ACHY:FMug}.

Some interesting papers \cite{GJ:McK,Tomas:McK,Mayr:McK} have appeared
recently which compute the finite monodromy associated to orbifold
theories by using the method of ``helices and mutations of exceptional
sheaves''. One of the motivations of this paper was to better
understand this construction in the language of the derived
category. We discuss the connection (and differences) in section
\ref{s:hel}. We also study the 
case of reducible exceptional divisors in section \ref{ss:red} which
appears, at least at first sight, to lie somewhat outside the method of
helices.

At least in the context of Batyrev-type \CY\ varieties associated to
toric geometry \cite{Bat:m}, it seems possible to rigorously classify
all types of monodromy. Indeed Horja \cite{Horj:DX} has achieved this
in the neighbourhood of the large radius limit. Rather than attempt
such a classification we will simply go through some examples which
appear to demonstrate most of the interesting things which can happen.

Our analysis is closely tied to the ``phase picture''
\cite{W:phase,AGM:II} of the moduli space. One has various limit
points in the moduli space each of which lies in the centre of some
phase. There are naturally embedded $\P^1$'s in the moduli space which
connect adjacent limit points. We are concerned with monodromy within,
or almost within, such $\P^1$'s. Most of the ``interesting'' questions
one could ask about monodromy appear to be contained in this structure.

In section \ref{s:two} we exhaustively study a two-parameter example
obtaining all interesting monodromies associated to this model. In
section \ref{s:other} we study some aspects of another couple of
examples which exhibit some properties not seen in section \ref{s:two}.

Because of the prominent r\^ole played by the 0-brane in the
construction of Bondal and Orlov, we discuss some of its properties
under monodromies in section \ref{s:D0}.

%%%%%%%%%%%%%%%%%%%%%%%%%%%%%%%%%%%%%%%%%%%%%%%%%%%%%%%%%%%%%%%%%%%

\section{Autoequivalences and the Fourier-Mukai Transform}
  \label{s:FMT}

In this section we will quickly review the language we use for the
derived category. See \cite{GM:Hom,Thom:DCg} for more information
about the derived category itself and \cite{ST:braid,Thom:DX-l} for
more on some of the notation used below.

Douglas \cite{Doug:DC} has argued that the even-dimensional D-branes
on a \CY\ $X$ should be associated with objects in the derived category of
coherent sheaves on $X$. The morphisms in this category are associated
to open strings. Given a particular object $\cK$ of $\DC(X\times X)$ we
associate projections
\begin{equation}
\xymatrix{
  &X\times X\ar[dl]_{p_1}\ar[dr]^{p_2}\\
  X&&X
}
\end{equation}
and the Fourier--Mukai transform \cite{Muk:FM,Muk:FM2}
\begin{equation}
  T_{\cK}(\cF)=\mathbf{R}p_{2*}(\cK\Lotimes p_1^*\cF),  \label{eq:FM}
\end{equation}
for any object $\cF$ of $\DC(X)$.

If $\cK$ is chosen carefully (see \cite{BO:DCeq,Brid:faith} for
details) then the Fourier--Mukai transform will be an
``autoequivalence'' of $\DC(X)$. Namely it maps $\DC(X)$ back to
itself while preserving the important algebraic structure associated
to the ``distinguished triangles''. What this means for us is that the
physics should remain invariant under such a transformation.

There are two cases of such $\cK$'s which are of particular
interest which will be denoted $\cK^B$ and $\cK^K$. First let $L$ be a
line bundle (or invertible sheaf) over 
$X$ and let $j:X\to X\times X$ be the diagonal embedding. Then let
$\cK^B_L$ (where the superscript $B$ stands for ``$B$-field'' for
reasons to become clear) be the object of $\DC(X\times X)$ given by 
\begin{equation}
  \ldots\to0\to j_* L\to0\to\ldots,
\end{equation}
such that the nontrivial term is at the 0th position.

Now consider the object of $\DC(X)$ given by a sheaf at 0th
position:
\begin{equation}
  \ldots\to0\to \mathcal{F}\to0\to\ldots \label{eq:Fsheaf}
\end{equation}
If we apply to (\ref{eq:Fsheaf}) the Fourier--Mukai transform associated
to $\cK^B_L$, we obtain
\begin{equation}
  \ldots\to0\to \mathcal{F}\otimes L\to0\to\ldots
\end{equation}

To relate this to string theory, let $\mathcal{F}$ be a sheaf
supported over some subspace of $X$. That is, we have a D-brane
wrapping this subspace.
All we have done by applying this transform is to change the field-strength
of the $\GU(1)$ gauge bundle over the D-brane.
Gauge invariance forces the combination
$F-B$ to be appear in the action of the D-brane. The above
transformation must therefore be 
equivalent to $B\mapsto B+L$. (Note that we use $L$ to denote the
line bundle, the associated divisor class and the dual 2-form $c_1(L)$.) Thus
we may assert (as was also done in \cite{Horj:DX}) that {\em 
the transformations associated to $\cK^B_L$ are those of a 
$B$-field shift $B\mapsto B+L$.}

Another transformation of interest is that of Seidel and Thomas
\cite{ST:braid} given by
\begin{equation}
  \cK^K_{\cE} = \textrm{Cone}\Bigl\{\cE^\vee\boxtimes
  \cE\to j_*\O_X\Bigr\},   \label{eq:FMST}
\end{equation}
where $\O_X$ is the structure sheaf of $X$. The superscript $K$ stands
for ``Kontsevich'' who was the first to use this kind of transformation in the
context of string theory \cite{Kont:mon}.
Here the notation $A\boxtimes B$ is short for $p_1^*A\otimes
p_2^*B$. The object $\cE$ is any object of $\DC(X)$ which satisfies the
sphericity conditions given in \cite{ST:braid}. We refer to
\cite{Thom:DCg} for a nice description of the cone construction.

Now the associated Fourier--Mukai transform simplifies to the following
\cite{ST:braid}: 
\begin{equation}
  T_{\cK^K_{\cE}}(\cF) = \textrm{Cone}\Bigl\{\Hom(\cE,\cF)
  \otimes\cE \labto{f} \cF\Bigr\},  \label{eq:ST2}
\end{equation}
where $f$ is the obvious evaluation map.

It is worth pointing out a subtle but potentially important point. The
cone construction is not a particularly well-defined functor in the
context of the derived category. We refer to section 1.4 of chapter 5
of \cite{GM:Hom} for a discussion of the problems. By writing the
transformation in the form (\ref{eq:ST2}) we potentially expose
ourselves to such ambiguities. One should always bare in mind however
that the transformation exists as a Fourier--Mukai transform (\ref{eq:FM})
which yields a perfectly well defined functor from $\DC(X)$ to itself.
In particular the cone appearing in (\ref{eq:FMST}) is only defining 
$\cK^K_{\cE}$ as an object and no functorial properties of
the cone are required there.

It is difficult to understand the physical meaning of the
transformation associated to $\cK^K_{\cE}$ working directly in the
derived category. Instead we take Chern 
characters to see the effect on cohomology.
From (\ref{eq:ST2}) one deduces that \cite{ST:braid}
\begin{equation}
  \ch(T_{\cK^K_{\cE}}(\cF)) = \ch(\cF) -
  \langle\cE,\cF\rangle\ch(\cE), \label{eq:PL} 
\end{equation}
where 
\begin{equation}
\begin{split}
  \langle\cE,\cF\rangle &= \sum_i (-1)^i\dim\Hom^i(\cE,\cF)\\
     &= \int_X\ch(\cE^\vee)\ch(\cF)\td(\cT_X),
\end{split}
\end{equation}
and $\cT_X$ is the tangent sheaf of $X$.

Now it is generally believed that the (skew-symmetric) inner product
$\langle\cE,\cF\rangle$ on $X$ is equal to the (equally
skew-symmetric) intersection form for 3-cycles on the mirror $Y$
\cite{HM:alg2,BDLR:Dq,HIV:D-mir}.
According to this analogy, the transformation (\ref{eq:PL}) is nothing
more than a Picard--Lefschetz transformation that one would associate
to monodromy around a vanishing 3-sphere in $Y$
\cite{ST:braid,HIV:D-mir}. Because of this it seems natural to
expect this kind of transformation to be associated to monodromy in the
moduli space of complex structures around some parts of the
discriminant locus.

%%%%%%%%%%%%%%%%%%%%%%%%%%%%%%%%%%%%%%%%%%%%%%%%%%%%%%%%%%%%%%%%%%%

\section{A One Parameter Case}   \label{s:one}

In this section we will review a computation apparently first done by
Kontsevich \cite{Kont:mon}.

We consider the case where $X$ is the quintic hypersurface in
$\P^4$. As is very well-known \cite{CDGP:} the moduli space of
complexified K\"ahler forms can be taken to be $\P^1$ with three
interesting point. These point are as follows:
\begin{itemize}
  \item[$P_0$:] The Gepner point. Metrically it lies at an orbifold
  point $\C/\Z_5$.
  \item[$P_1$:] The ``conifold point''. The mirror of $X$ acquires a
  conifold singularity. The conformal field theory
  associated to $X$ is singular.
  \item[$P_\infty$:] The large radius limit. This point is an infinite
  distance away from the above two points.
\end{itemize}

Let $H$ denote the homology class of the 4-cycle given by the
hyperplane section of the quintic 3-fold. We will also use $H$ to
denote its Poincar\'e dual which generates $H^2(X,\Z)$. We then have
\begin{equation}
\begin{split}
  \td(\cT_X) &= \frac{\left(\displaystyle\frac{H}{1-e^{-H}}\right)^5}
	{\displaystyle\left(\frac{5H}{1-e^{-5H}}\right)}\\
  &= 1 + \ff56H^2,
\end{split}
\end{equation}
and
\begin{equation}
  \int_X H^3=5.
\end{equation}

The monodromy around $P_0$ is expected to be of order 5 because of the
$\Z_5$ quantum symmetry of the Gepner model. The monodromy around
$P_\infty$ is known to correspond to $B\mapsto B+H$. In other words we
expect it to be given by the transformation $\cK^B_{\O(H)}$. We will
denote this by $\cK^B_H$ for short.

Kontsevich conjectured that the monodromy around $P_1$ is given by the
Fourier--Mukai transform $\cK^K_\cE$ of the previous section where $\cE$
is given by the structure sheaf $\O_X$. We denote this $\cK_0^K$ for short.
We will return to this conjecture in a more precise form in section
\ref{s:two}. 

It follows from the topology of a sphere with 3 punctures that the
product\footnote{Note that we are required to take the loops around
$P_1$ and $P_\infty$ in the ``same direction'' in order for their
product to be a loop around $P_0$. Throughout this paper we will have
orientation problems such as this. We will not concern ourselves at
all with such details. In all the examples we do, we simply find the
right combination which gives the expected results.} of the monodromy
around $P_\infty$ and the monodromy around $P_1$ should equal the
monodromy around $P_0$ and hence should be of order 5. We may verify
that this is consistent with the Chern character of any starting
D-brane.

Let us start with $\cF$ given by $\O_X$ and apply the desired sequence
of monodromy transformations. From (\ref{eq:PL}) we obtain the
following:
\begin{equation}
\begin{split}
\ch(\cF) &= 1\\
\ch(\cK^B_H\cF) &= e^H\\
\ch(\cK^K_0\cK^B_H\cF) &= e^H-5\\
\ch(\cK^B_H\cK^K_0\cK^B_H\cF) &= e^{2H}-5e^H\\
\vdots&\\
\ch((\cK^K_0\cK^B_H)^5\cF)&=1+(e^H-1)^5=1,
\end{split}
\end{equation}
as $H^4=0$ in $X$. This is therefore consistent.

It would be interesting to check that $(\cK^K_0\cK^B_H)^5$ gives the
identity transform when applied directly to $\DC(X)$ rather than just
applied to cohomology. We will not attempt this here.

Note that it is easy to apply the same method to other one
parameter examples as listed in \cite{Sch:12mon} for example.

%%%%%%%%%%%%%%%%%%%%%%%%%%%%%%%%%%%%%%%%%%%%%%%%%%%%%%%%%%%%%%%%%%%

\section{Helices and Mutations} \label{s:hel}

The purpose of this section is to briefly point out similarities and
differences between the above computation for the quintic and the
notion of helices and exceptional sheaves. The reader
who is not directly interested in such things can skip this section.

\def\clE{\cE}
It is easiest to describe mutations and helices directly in the
derived category. See \cite{Gorod:Dhel} for example for more
information.
Let us consider the space $V=\P^4$ and let $H$ denote the hyperplane
class. Now consider the exceptional collection of sheaves
$\{\O,\O(H)\}$. We may use a mutation to pull $\O(H)$ to the left
through $\O$. Let $\clE(H)$ denote the resulting object in
$\DC(\P^4)$. One can compute $\ch(\clE(H))=e^H-5$. 

Similarly we may begin with the set $\{\O,\O(H),\O(2H)\}$ and pull
$\O(2H)$ through $\O(H)$ and $\O$ to obtain $\clE(2H)$ etc. The result of
such mutations appears remarkably similar to the monodromy
transformations we considered in the previous section. Indeed one
obtains
\begin{equation}
  \ch(\clE(nH)) = \ch((\cK^K_0\cK^B_H)^n\cF),\quad\text{for
  $n=0,\ldots,4$.}   \label{eq:mut}
\end{equation}

This correspondence fails for $n=5$ however. In this case we have 
$\ch(\clE(5H))=0$. This disagreement should come as no surprise. The
language of mutations of helices can be recast in the form of the
Fourier--Mukai transforms of section~\ref{s:FMT}. The key point
however is that the algebraic variety in question is the ambient
$\P^4$ itself rather than the \CY\ hypersurface $X$. Indeed
exceptional sheaves cannot exist on $X$.

A Fourier--Mukai transformation does {\em not\/} yield an autoequivalence
of $\DC(\P^4)$ as it fails the canonical class constraint of 
\cite{Brid:faith}. That is why the fifth application of the supposed
monodromy transformation can kill the object in $\DC(\P^4)$.

The language of mutations of helices was used in
\cite{GJ:McK,Tomas:McK,Mayr:McK} successfully to obtain monodromies
because equation (\ref{eq:mut}) holds true. Note however that the
procedure of going from $\clE(nH)$ to $\clE((n+1)H)$ cannot generally
be identified with monodromy around the Gepner point because
of the more general failure of this relation.

%%%%%%%%%%%%%%%%%%%%%%%%%%%%%%%%%%%%%%%%%%%%%%%%%%%%%%%%%%%%%%%%%%%

\section{A Two Parameter Case}   \label{s:two}

The structure of monodromies becomes considerably more interesting
when one starts to look at moduli spaces of more than one
dimension. In this case, the ``discriminant locus'' of bad conformal
field theories is a subvariety of the moduli space with dimension one
or more.

We wish to see if the Fourier--Mukai transforms considered above can
also be applied in these more complicated situations. Note that this
problem has been studied by Horja \cite{Horj:DX} and by  Seidel and
Thomas \cite{ST:braid}. Horja gave
extensive results for monodromy loops which are ``close'' to the large
radius limit in some sense. We will be interested in relations
obtained by venturing further into the moduli space. The
Fourier--Mukai transforms we will consider follow closely the
construction by  Seidel and Thomas \cite{ST:braid}. The paper
\cite{Horj:EZ} has appeared very recently which 
shows that these methods are essentially a special case of Horja's
construction.

Our goal in this section will be to obtain a similar result to section
\ref{s:one} in a two parameter example. Namely, can we find a sequence
of monodromies which give a loop around some point which looks like a
Gepner point, and hence has finite order? 

Because the moduli space is two-dimensional there is no notion of
``monodromy around a point''. Given a complex curve in our moduli
space we can define a monodromy. We will use two notions extensively
in the text below 
and we wish to emphasize the difference here to avoid confusion. We
will often refer to monodromy {\em around\/} a curve for a loop in the
two-dimensional moduli space which lies external to the curve.  We
will also refer to to the completely different notion of monodromy {\em 
within\/} the curve around a specific point. The words ``around'' and
``within'' will always have the above meaning.

Our example is where $X$ is a blown-up hypersurface of degree 18 in the
weighted projected space $\P^4_{9,6,1,1,1}$. This space was studied
extensively in \cite{CFKM:II} and analyzed in relation to D-branes in
\cite{DR:Dell}. The two generators of $H^2(X,\Z)$ (and 
their Poincar\'e dual divisors) will
be called $H$ and $L$ consistent with \cite{CFKM:II}. If we put
homogeneous coordinates $[z_1,\ldots,z_5]$ on $\P^4_{9,6,1,1,1}$ then
the divisor class of $z_1=0$ is given by $3H$, the class of $z_2=0$ is
given by $2H$. The class of $z_3=0$ or $z_4=0$ or $z_5=0$ (after the
blow-up) is given by $L$.
This weighted projective space has a curve of singularities which
intersects the hypersurface at one point. Locally this point looks like
the orbifold $\C^3/\Z_3$. We blow this up to produce an exceptional
divisor $E\cong\P^2$. In terms of homology classes, $E=H-3L$.

\iffigs
\begin{figure}
  \centerline{\epsfxsize=10cm\epsfbox{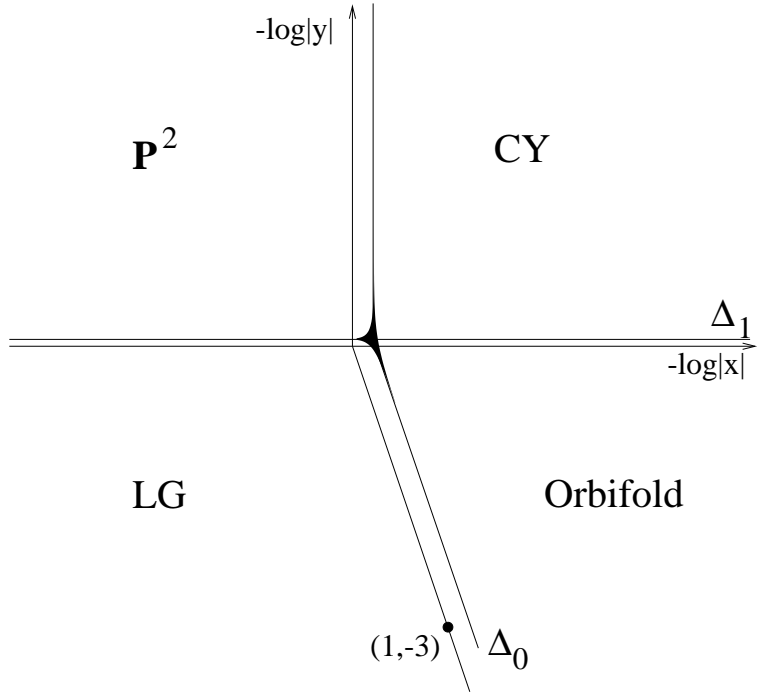}}
  \caption{The 4 phases of the 2 parameter example.}
  \label{fig:phase}
\end{figure}
\fi

The remaining topological information for $X$ required is as
follows. $H^2$, $HL$, $L^2$ live in $H_2(X)$ or $H^4(X)$ subject to the
constraint $H(H-3L)=0$. $H_0(X)$ or $H^6(X)$ has a single generator we
denote $p$, and $H^3=9p$, $H^2L=3p$, $HL^2=p$, $L^3=0$. 
Obviously any monomial of degree 4 or higher in $H$ or $L$ vanishes.
Finally
\begin{equation}
  \td(\cT_X) = 1-\ff12L^2 +\ff14HL.
\end{equation}

The mirror, $Y$, of $X$ has defining equation
\begin{equation}
  a_0z_1z_2z_3z_4z_5 + a_1z_1^2 + a_2z_2^3 + a_3z_3^{18} + a_4z_4^{18}
  + a_5z_5^{18} + a_6z_3^6z_4^6z_5^6.  \label{eq:def1}
\end{equation}
The ``algebraic'' coordinates on the moduli space are then given by
\begin{equation}
  x=\frac{a_1^3a_2^2a_6}{a_0^6},\quad y=\frac{a_3a_4a_5}{a_6^3}.
	\label{eq:algco}
\end{equation}

We may then define the discriminant as an expression in $x$ and $y$
which vanishes when $Y$ becomes singular. If the data can be presented
torically as in this case then section 3.5 of \cite{MP:inst} gives a
nice efficient way of computing this discriminant. In our example, the
discriminant factorizes into two parts:
\begin{equation}
\begin{split}
  \Delta_0 &= 6^{12}x^3y + (432x-1)^3\\
  \Delta_1 &= 27y+1.
\end{split}
\end{equation}

We want to picture the moduli space in two different ways. First we
use the ``phase'' description of \cite{W:phase,AGM:II}. See also
section 3 of \cite{AGM:sd}. We
project the discriminant into $\R^2$ by plotting $-\log|y|$ against
$-\log|x|$. We show the result in figure~\ref{fig:phase}. The result
is that the 2-plane is divided into four ``phases''. We chose our
algebraic coordinates (\ref{eq:algco}) so that the \CY\ phase appears
in the positive quadrant. The limit point of this phase is the large
radius limit.
The positive quadrant may also be viewed as the
K\"ahler cone of $X$ where the class $H$ gives the horizontal
direction and $L$ gives the vertical direction.

The other phases are pictured as follows. There is an orbifold phase
whose limit point has the orbifold singularity $\C^3/\Z_3$ but the
\CY\ has infinite volume. There is a $\P^2$ phase where $X$ collapses
onto a $\P^2$. This can happen as $X$ is an elliptic fibration over
$\P^2$. In the limit, this elliptic fibre has zero area and the $\P^2$
has infinite volume. Finally we
have a Landau--Ginzburg phase with the Gepner point as the limit point.

The other way of drawing the moduli space is as a complex surface. The
phase picture in figure~\ref{fig:phase}, i.e., the {\em secondary
fan\/} of $X$, is viewed as the fan of a toric variety $\cM$ as
described in \cite{AGM:II}. 
The coordinates $x$ and $y$ are then naturally coordinates over a
patch of the moduli space $\cM$.\footnote{This is identified by associating the
dual of the cone in the positive quadrant with $\Spec \C[x,y]$.} In this
case $\cM$ is a surface with two quotient singularities.  The
discriminant is a divisor in $\cM$.  

\iffigs
\begin{figure}
  \centerline{\epsfxsize=10cm\epsfbox{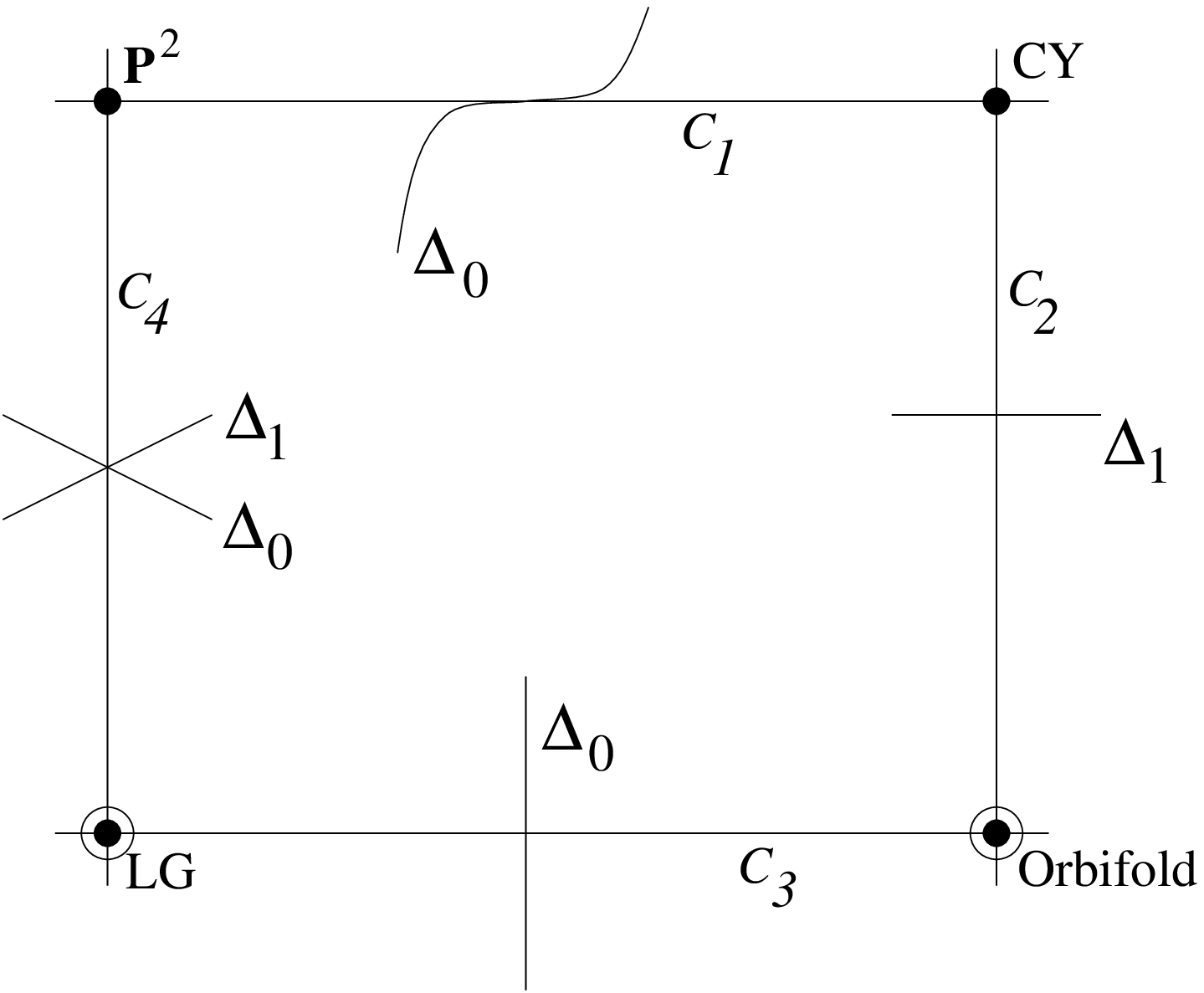}}
  \caption{The moduli space $\cM$.}
  \label{fig:M}
\end{figure}
\fi

We sketch $\cM$ in figure~\ref{fig:M} by drawing complex dimensions as
real. Our limit points appear as dots
in the diagram. We draw the curves $C_1,\ldots,C_4$ as the $\P^1$'s
joining adjacent phase limits. Torically these curves are associated
to the lines in figure \ref{fig:phase} which separate the
phases. Figure~\ref{fig:M} also shows how the discriminant 
intersects these curves. Note that $\Delta_0$ and $\Delta_1$ are
themselves smooth curves in $\cM$. The extra circles around the LG and
orbifold point denote the fact that both of these points lie at a
quotient singularity of the form $\C^2/\Z_3$ in $\cM$.

If we followed the analysis of \cite{CFKM:II} we would now blow-up
$\cM$ so that it was smooth and that the intersections of the
discriminant with the curves $C_i$ were transverse. This
requires several blow-ups. Rather than do this, we find it easier to
work directly in $\cM$.

We would now like to take each of the curves $C_1,\ldots,C_4$ in turn
and do a similar computation to that of section \ref{s:one} {\em
within\/} that curve (or nearly so) to check
our monodromy predictions. Note that each curve $C_i$ has three
special points on it --- the two limit points and a third point where
the discriminant hits the curve in some way. Thus, just as in section
\ref{s:one}, we will show that the product of the monodromy around one
of the limit points and around the discriminant is equal to the
monodromy around the other limit point.

\subsection{$C_1$}  \label{ss:C1}

Let us fix a basepoint near the large \CY\ limit point. Because we
have identified the cone of this phase with the K\"ahler cone of $X$
we immediately know the monodromies around $C_1$ and $C_2$. Each must
be a shift in the $B$-field. To be precise, a loop around $C_1$ will
correspond to $B\mapsto B+L$ and hence corresponds to
$\cK^B_L$. Similarly a loop around $C_2$ is given by $\cK^B_H$.

Since $C_1$ and $C_2$ intersect transversely, the monodromy around the
\CY\ point within $C_1$ corresponds to going around $C_2$ and is thus given by
$\cK^B_H$. In other words $\ch(\cF)\mapsto e^H\ch(\cF)$.

What about the monodromy within $C_1$ around the point in the middle
where the discriminant hits? A method for computing this was presented
in \cite{Horj:DX} but we will proceed a little differently. First we need to
decide how to go around a generic part of $\Delta_0$. Note that
$\Delta_0$ is the irreducible component of the discriminant
corresponding to the appearance of singularities in (\ref{eq:def1})
for {\em nonzero\/} $z_1,\ldots,z_5$. This was dubbed the
``$A$-discriminant'' in \cite{GKZ:book}.\footnote{Perhaps rather
confusingly, \cite{GKZ:book} uses the term ``principal
$A$-determinant'' for the full discriminant $\Delta_0\Delta_1$. Even
more confusingly, 
$\Delta_0$ has sometimes been called the ``principal component'' of the
discriminant \cite{Horj:DX}.}
We will call it the ``{\em primary\/}'' component of the discriminant.
One could also define this as the component which is
computed by finding solutions to equation (3.45) of \cite{MP:inst}. We
then state the following conjecture which appears to be due to
Horja, Kontsevich and Morrison in some form or another
\cite{Kont:mon,Mor:geom2,Horj:DX}.
\begin{conjecture}
For a suitable choice of basepoint near the \CY\ limit point, a loop
around the primary component of the discriminant is given by
$\cK_0^K=\cK^K_{\cE}$, with $\cE=\O_X$.
  \label{conj:K}
\end{conjecture}
This is certainly consistent with section \ref{s:one} where the primary
component was the entire discriminant.

Assuming this conjecture to be true we still have a complication that
makes the computation a little less straight-forward. Namely, $C_1$
does not intersect $\Delta_0$ transversely but rather intersects it
tangentially with multiplicity 3. This means that we cannot say that
the monodromy within $C_1$ around the discriminant is given by the
above conjecture.

\iffigs
\begin{figure}
  \centerline{\epsfxsize=7cm\epsfbox{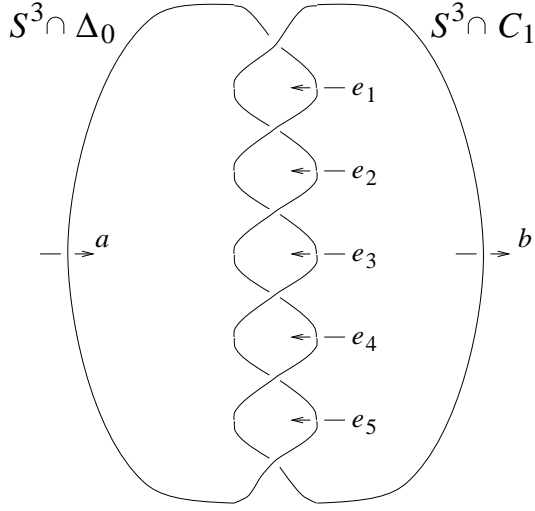}}
  \caption{The triply linked circles.}
  \label{fig:3link}
\end{figure}
\fi

To proceed with the computation we can put a small 3-sphere $S^3$ around the
intersection of $\Delta_0$ and $C_1$. Since $\Delta_0$ and $C_1$ are
both smooth it follows that $L_1=S^3\cap C_1$ and
$L_2=S^3\cap\Delta_0$ are both unknotted circles. Because of the
tangential intersection, these circles are linked three times. We may
remove a point of $S^3$ and imagine the link in $\R^3$. We show
this in figure~\ref{fig:3link}. 

Next we need to describe $\pi_1(S^3-(L_1\cup L_2))$. To do this we use
the Wirtinger presentation (see \cite{Rolf:knot} for example). Imagine
fixing a basepoint above the sheet of paper. The arrow $a$ in
figure~\ref{fig:3link} then represents the element of $\pi_1$ looping
under the left circle in the direction indicated by the
arrow. Similarly we define $b$ for the right circle. We then have
further elements $e_i$ as shown in the figure. Crossing relations then
determine\footnote{Our order convention is that $ab$ represents the path $b$
followed by the path $a$.}
\begin{equation}
\begin{split}
  e_1 &= b^{-1}ab\\
  e_2 &= b^{-1}a^{-1}bab\\
  e_3 &= b^{-1}a^{-1}b^{-1}abab\\
  e_4 &= b^{-1}a^{-1}b^{-1}a^{-1}babab\\
  e_5 &= b^{-1}a^{-1}b^{-1}a^{-1}b^{-1}ababab,
\end{split}
\end{equation}
but clearly $e_5=a$ which yields the relation
\begin{equation}
ababab=bababa.   \label{eq:pi1rel}
\end{equation}
Indeed $\pi_1(S^3-(L_1\cup L_2))$ is given by the group on two
generators $(a,b)$ subject to the single relation (\ref{eq:pi1rel}).

Deform the path within $C_1$ around the discriminant point a little so
that it lies outside $C_1$. This is the path around which we wish to
compute the monodromy.  This is the (clockwise) path which follows
closely the circle $S^3\cap C_1$. Such a path is homotopic to
$e_1e_3e_5= b^{-2}ababa$. But $a$ is nothing more than a generic loop
around $\Delta_0$ and so is given by $\cK_0^K$, and similarly $b$ is
given by $\cK^B_L$. Therefore we claim that monodromy around the
discriminant point within $C_1$ is given by
$(\cK^B_L)^{-2}\cK_0^K\cK^B_L\cK_0^K\cK^B_L \cK_0^K$.

Let $Q_1$ denote the monodromy within $\C_1$ around the $\P^2$ limit
point. As this is equal to the combined monodromy around the
discriminant point and the \CY\ limit point, we have
$Q_1=\cK^B_H(\cK^B_L)^{-2}\cK_0^K\cK^B_L \cK_0^K\cK^B_L\cK_0^K$.

Passing from the \CY\ limit point to the $\P^2$ limit point
represents collapsing a large radius elliptic fibre to a
Landau--Ginzburg orbifold theory in a manner very similar to the
collapse of the quintic in section \ref{s:one}. This Landau-Ginzburg
theory is a $\Z_6$-orbifold and thus has a $\Z_6$ quantum symmetry. It
should therefore follow that $Q_1^6=1$ in complete analogy with the
Landau--Ginzburg point in section \ref{s:one}. This is a highly
nontrivial check of our picture:
\begin{equation}
\begin{split}
\ch(\O_X) &= 1\\
\ch(Q_1\O_X) &= e^H - 3e^{H-L} + 3e^{H-2L}\\
\ch(Q_1^2\O_X) &= e^{2H} - 3e^{2H-L} + 3e^{2H-2L} -e^H\\
\vdots\\
\ch(Q_1^6\O_X) &= e^{6H} - 3e^{6H-L} + 3e^{6H-2L} -e^{5H} -e^{4H}
  +3e^{4H-L}-3e^{4H-2L} +3e^{3H-L}\\&\qquad-3e^{3H-2L}+e^{2H}
  +e^H-3e^{H-L}+3e^{H-2L}\\
  &= 1.
\end{split} \label{eq:Q1t}
\end{equation}

%%%%%%%%%%%%%%%%%

\subsection{$C_2$}  \label{ss:C2}

So far we have only used $\cE=\O_X$ in the Fourier--Mukai transform
$\cK^K_{\cE}$. In this section we use a less trivial choice.
The curve $C_2$ represents the process of blowing-up up the $\C^3/\Z_3$
singularity while keeping the rest of the \CY\ at infinite volume. We
will attempt to ``localize'' the computations to around the
exceptional divisor $E\cong\P^2$. 

Let us consider the general case of an irreducible exceptional divisor
$E$ in a \CY\ space $X$ of arbitrary dimension. Let the normal bundle
be denoted by $N$. Let us assume that the zero locus of a generic section
of $N^\vee$
gives an irreducible variety $W\subset E$ of complex dimension two less than
$X$. $W$ is automatically \CY. In our example, $W$ would be a cubic
curve in $\P^2$. We 
therefore have two inclusions
\begin{equation}
  i:E\hookrightarrow X,\quad j:W\hookrightarrow E.
\end{equation}
Now consider two objects $\cE,\cF\in\DC(E)$ associated to sheaves on
$E$. In terms of our inner product on $X$ we may apply the
Grothendieck--Riemann--Roch theorem to yield the following localization:
\begin{equation}
\begin{split}
  \langle i_*\cE,i_*\cF\rangle_X &= \int_X (\ch(i_*\cE))^\vee
	\ch(i_*\cF)\td(\cT_X)\\
  &= \int_E (\ch(i_*\cE))^\vee \ch(\cF)\td(\cT_E)\\
  &= \int_W (\ch(j^*\cE))^\vee \ch(j^*\cF)\frac{\td(\cT_E)}{\td(N^\vee)}\\
  &= \int_W (\ch(j^*\cE))^\vee \ch(j^*\cF)\td(\cT_W)\\
  &= \langle j^*\cE,j^*\cF\rangle_W.
\end{split}
\end{equation}
Therefore we may compute the inner product between objects of $\DC(X)$
which are $i_*$ of something in $\DC(E)$ purely in terms of the local
geometry of the blow-up.

We may now apply conjecture~\ref{conj:K} to $W$. In our example $W$ is
an elliptic curve and has only one deformation of complexified
K\"ahler form. The discriminant is then a point and therefore
primary with respect to $W$. The associated Fourier--Mukai
transform for monodromy is then given by $\cK^K_{\cE}$ for
$\cE=\O_W=j^*\O_E$.
This naturally motivates the following
\begin{conjecture}
The monodromy around a component of the discriminant associated with
an irreducible divisor $E$ collapsing to a point is given by $\cK^K_{\cE}$
for $\cE=i_*\O_E$, where $i$ is the inclusion map. 
\end{conjecture}

Note that $i_*\O_E$ is the structure sheaf of $E$ extended by zero
over the rest of $X$. It can thus be denoted by $\O_E$ itself.

In our example we therefore associate $\Delta_1$ with $\cK^K_{\cE}$
for $\cE=\O_E$ where the class of $E$ is given by $H-3L$. Let us use
$\cK^K_1$ to denote this transform. Another
application of Grothendieck--Riemann--Roch quickly yields
\begin{equation}
  \ch(\O_E) = 1-e^{3L-H},
\end{equation}
(and so $\ch(\O_E^\vee) = 1-e^{H-3L}$).

We are now in a position to compute all the monodromies for
$C_2$. Around the \CY\ limit we have $\cK^B_L$. The component
$\Delta_1$ hits $C_2$ transversely and so the monodromy around the
discriminant point is given by $\cK^K_1$.

Let us use $Q_2=\cK^K_1\cK^B_L$ to refer to monodromy around the
orbifold limit point. 
A $\C^3/\Z_3$ orbifold has a $\Z_3$ quantum symmetry so one might
na\"\i vely guess that $Q_2^3=1$. Instead we find
\begin{equation}
\begin{split}
 \ch(\O_X) &= 1\\
 \ch(Q_2\O_X) &= e^L\\
 \ch(Q_2^2\O_X) &= e^{2L}\\
 \ch(Q_2^3\O_X) &= e^H.
\end{split}  \label{eq:Q2}
\end{equation}
This suggests the relation $Q_2^3=\cK^B_H$. To see why this is so let
us examine more 
carefully the geometry of the moduli space near the orbifold limit
point. As mentioned earlier this point is actually locally the
singularity $\C^2/\Z_3$. We may therefore surround this limit point, not by
a sphere, but by a {\em lens space\/} $M=S^3/\Z_3$. Now $C_2$ and
$C_3$ are both smooth curves and therefore they intersect $M$ in
unknotted circles.

Consider the free $\Z_3$ quotient map $q:S^3\to M$. The intersection 
of $C_2$ and $C_3$ with $M$ both lift to single circles in $S^3$ under
$q^{-1}$. These circles are linked once and so $\pi_1$ of the
complement of these circles in $S^3$ is the abelian product $\Z\times\Z$.
Let $G$ denote $\pi_1$ of the complement of $C_2$ and $C_3$ in
$M$. Since $q$ is a normal cover we have
\begin{equation}
  1\to\Z\times\Z\to G\to\Z_3\to1.   \label{eq:ext3}
\end{equation}
A more detailed analysis of the geometry shows that $G\cong\Z\times\Z$
where there is a particular element $g_{\mathrm{orb}}\in G$ which
cannot be lifted to 
a loop in $S^3$ but such that $g_{\mathrm{orb}}^3$ lifts to a loop in
$S^3$ which 
loops around both $C_2$ and $C_3$.

To understand the monodromy we need to deform the loop ``inside''
$C_2$ around the orbifold point a little so that it doesn't intersect
$C_2$ or $C_3$. There is no unique way to do this. The homotopy class
of such a deformation is given by $g_{\mathrm{orb}}$ times an
arbitrary number of 
windings {\em around\/} $C_2$. One might argue that the most natural
lift is the reduce this extra winding around $C_2$ and say that the
desired loop is simply given by $g_{\mathrm{orb}}$.

Identifying $Q_2$ with $g_{\mathrm{orb}}$ it should then follow that
the monodromy 
$Q_2^3$ is given by a loop around $C_2$ followed by a loop around
$C_3$. We see that $Q_2^3=\cK^B_H$ is entirely consistent with this so
long as {\em the monodromy around $C_3$ is trivial.} We also see that
our natural deformation of the loop within $C_2$ is the correct one.

We have therefore understood the monodromy in (\ref{eq:Q2}) and argued
that the monodromy around $C_3$ is trivial.

For a true localization to the orbifold point we may put $H=0$. This
has the effect of sending that component of the K\"ahler form of to
infinity. Thus the target space looks like a resolution of
$\C^3/\Z_3$. In this case the monodromy around the orbifold point
really is of order three. This fact was also determined directly using the
Picard--Fuchs system in \cite{DG:fracM}. Note also that any object of
$\DC(X)$ which can be written as $i_*$ of something in $\DC(E)$ is
brought back to itself exactly after looping three times around the
orbifold point.

%%%%%%%%%%%%%%%%%

\subsection{$C_3$}  \label{ss:C3}

Consider first the loop around the orbifold point within $C_3$. In
order to understand the monodromy we need again to deform this loop a
little as in section~\ref{ss:C2}. It turns out that the simplest
deformation of this loop is homotopic to the same class
$g_{\mathrm{orb}}$ as above. Figure~\ref{fig:M} certainly makes
this statement counterintuitive! At first sight it looks like a loop
within $C_2$ is like a loop around $C_3$ and a loop within $C_3$
looks like a loop around $C_2$ and these are certainly not equal. It
is the quotient singularity which stops this argument working. Both
the loop within $C_2$ and the loop within $C_3$ must deform to
elements of $G$ which map to the same element of $\Z_3$ in
(\ref{eq:ext3}). The most natural deformation of these two loops
actually makes the loops homotopic.

Therefore monodromy around the orbifold point within $C_3$ is given by 
$Q_2=\cK_1^K\cK_L^B$. The loop around the discriminant point within
$C_3$ is easy. Since $\Delta_0$ intersects $C_3$ transversely, the
monodromy is given by $\cK_0$. The monodromy around the LG point is
then given by
\begin{equation}
  Q_3 = \cK^K_0\cK^K_1\cK^B_L.
\end{equation}

What properties should we expect for $Q_3$? The geometry around the LG
point is very similar (up to orientation questions) to the geometry
around the orbifold point. In particular one may show that the loop
corresponding to $Q_3$ is such that its third power is homotopic to a
loop around $C_3$ and $C_4$. Now we know that a loop around $C_3$
induces no monodromy from section \ref{ss:C2}. The fact that $C_4$
intersects $C_1$ transversely at a smooth point in $\cM$ tells us that
the loop around $C_4$ is given by $Q_1$ from section \ref{ss:C1}. Thus 
$Q_3^3$ should have the same properties as $Q_1$. We saw in
section \ref{ss:C1} that $Q_1$ was of 
order 6. {\em It follows that $Q_3$ is of order 18.}

We may confirm this explicitly. E.g.:
\begin{equation}
\begin{split}
 \ch(\O_X) &= 1\\
 \ch(Q_3\O_X) &= e^L-3\\
 \ch(Q_3^2\O_X) &= e^{2L}-3e^L+3\\
 \ch(Q_3^3\O_X) &= e^{H}-3e^{2L}+3e^L-1\\
 \ch(Q_3^4\O_X) &= e^{H+L}-3e^{H}+3e^{2L}-e^L\\
 \ch(Q_3^5\O_X) &= e^{H+2L}-3e^{H+L}+3e^{H}-e^{2L}\\
 \ch(Q_3^6\O_X) &= e^{2H}-3e^{H+2L}+3e^{H+L}-e^{H}-1\\
 \ch(Q_3^7\O_X) &= e^{2H+L}-3e^{2H}+3e^{H+2L}-e^{H+L}-e^L+3\\
 \vdots\\
 \ch(Q_3^{18}\O_X) &= e^{6H} - 3e^{5H+2L} + 3e^{5H+L} -e^{5H} -e^{4H}
  +3e^{3H+2L}-3e^{3H+L} +3e^{2H+2L}\\&\qquad-3e^{2H+L}+e^{2H}
  +e^H-3e^{2L}+3e^{L}\\
  &= 1.
\end{split}  \label{eq:Q3t}
\end{equation}
Comparing closely (\ref{eq:Q1t}) and (\ref{eq:Q3t}) we see that
$Q_3^3$ is not quite the same thing as $Q_1$ although their effect is
very similar. As we have described the loops corresponding to these
transformations, they actually differ by a change in basepoint and so
$Q_1$ and $Q_3^3$ are only equivalent up to conjugation.

The sequence of transformations given in (\ref{eq:Q3t}) is identical
to the sequence predicted by \cite{GJ:McK,Tomas:McK,Mayr:McK} in the
language of helices and mutations. There one begins with a sequence of
exceptional sheaves on $\P^4_{\{9,6,1,1,1\}}$ of the form $\{\O,\O(L),
\O(2L),\O(H),\O(H+L),\O(H+2L),\O(2H),\ldots,\O(5H+2L)\}$. One then
mutates all the bundles to the left to reverse their order giving a
sequence of objects whose Chern characters are exactly given by
(\ref{eq:Q3t}). It is not hard to see why this is so. Roughly speaking
the transformation $Q_3=\cK^K_0\cK^K_1\cK^B_L$ may be described as
follows. $\cK^B_L$ takes the sheaf $\O(nL)$ to $\O((n+1)L)$. Next
$\cK^K_1$ leaves $\O(mH+L)$ or $\O(mH+2L)$ invariant but takes
$\O(mH+3L)$ to $\O((m+1)H))$. Finally $\cK^K_0$ is a ``left mutation''
just as it was for the quintic in section \ref{s:hel}.

This gives a natural explanation for the funny ``jump'' seen in the
required sequence of exceptional sheaves from $\O(mH+2L)$ to
$\O((m+1)H)$. It is effectively caused by the action of $\cK^K_1$.
Note again the following shortcoming of the method of using sheaves
on $\P^4_{\{9,6,1,1,1\}}$. If we extend this process by adding
$\O(6H)$ to the above sequence of sheaves then the 18th transformation
of $\O$ would have Chern character equal to 0 rather than 1. Again
this is because the corresponding Fourier--Mukai transform on
$\P^4_{\{9,6,1,1,1\}}$ is not invertible.

%%%%%%%%%%%%%%%%%

\subsection{$C_4$}  \label{ss:C4}

Finally we do the monodromy computation within $C_4$. This actually
yields nothing new. Let $Q_4$ be the monodromy given by the loop within
$C_4$ around the LG point. Because the LG point is an orbifold point,
one can show that this loop (or at least a small deformation of it) is
the same as the loop within $C_3$ around the LG point for reasons
essentially identical to the discussion in section
\ref{ss:C2}. This immediately implies $Q_4=Q_3$.

Indeed monodromy around the $\P^2$ point within $C_4$ is given by
$\cK^B_L$; and monodromy around the discriminant point requires loops
around both $\Delta_1$ and $\Delta_0$ as seen in figure
\ref{fig:M}. Thus we see $Q_4=\cK^K_0\cK^K_1\cK^B_L$ consistent with
the above paragraph. 

Consider trying to generalize the results of this example.
One can show that the specific structure of the monodromy seen in this
example depends mainly on the fact that $X$ is a fibration. For
example we could consider a more 
complicated example with 3 K\"ahler moduli such as the resolved
hypersurface of degree 24 in $\P^4_{\{1,1,2,8,12\}}$. This is an
elliptic fibration over a Hirzebruch surface which itself is a
$\P^1$-fibration over $\P^1$. In this example one has a curve in the
moduli space which is the analogue of $C_4$ above. It connects the
Landau--Ginzburg phase with the $\P^1$ phase. Monodromy around the LG
point can then be shown to be of a form
$\cK^K_0\cK^K_1\cK^K_2\cK^B_L$. This will then reproduce the results
of section 9.3 of \cite{Mayr:McK} for example. Note however that the
fibration structure is essential here.

%%%%%%%%%%%%%%%%%%%%%%%%%%%%%%%%%%%%%%%%%%%%%%%%%%%%%%%%%%%%%%%%%%%

\section{Other Examples}   \label{s:other}

The example of section \ref{s:two} demonstrated many features of
monodromy but there are many other important possibilities which did
not appear. In this section we discuss some examples which do exhibit
these effects.

%%%%%%%%%%%%%%%%%%%%
\subsection{Surfaces shrinking to curves}  \label{ss:StoC}

All the monodromies around components of the discriminant have been
given by a Fourier--Mukai transform of the form (\ref{eq:FMST})
studied by Seidel and Thomas. Here we discuss an example that falls
outside this class.

Let us consider the resolved hypersurface $X$ of degree 8 in
$\P^4_{\{2,2,2,1,1\}}$. We refer to \cite{CDFKM:I} for extensive
details of this model. The space $X$ can be thought
of as a K3-fibration over $\P_1$ or as the resolution of a singular
space with a curve of singularities of the form $\C^2/\Z_2$. This
model has the same moduli space as that of section
\ref{s:two} except for the following aspects:
\begin{itemize}
  \item The $\P^2$ phase is renamed a $\P^1$ phase as this is the base
  of the fibration.
  \item The LG and orbifold limit points are now at orbifold points locally
  of the form $\C^2/\Z_2$.
  \item The $\Delta_0$ component of the discriminant now intersects
  $C_1$ at a point of multiplicity two rather than three.
\end{itemize}

When we compute the monodromies around the discriminant the real
difference appears when we consider $\Delta_1$.
Let $E$ be the exceptional divisor in $X$ coming from the resolution
of the curve of singularities. $E$ is a product of a genus 3 curve $Z$ and
$\Gamma\cong\P^1$. We associate $\Delta_1$ with the collapse of $E$
down to $Z$. The discussion in section \ref{ss:C2} concerned an
exceptional divisor contracting to a {\em point\/} and so cannot be
applied to monodromy around $\Delta_1$.

The computation of the monodromy on Chern characters was computed in
\cite{KMP:enhg}. The result may be rephrased as follows. For a divisor
$E$ collapsing to a curve $Z$ of genus $g$, monodromy around
$\Delta_1$ is given by
\begin{equation}
\ch(\cF) \mapsto \ch(\cF)-\langle\O_E+(1-g)\O_\Gamma,\cF\rangle
  \ch(\O_\Gamma) + \langle\O_\Gamma,\cF\rangle\ch(\O_E),
	\label{eq:EtoZ}
\end{equation}
where $\Gamma\cong\P^1$ is the inverse image of a point for the
blow-down.

The Fourier--Mukai transform given by (\ref{eq:FMST}) is incompatible
with (\ref{eq:EtoZ}). A more general form which is consistent is given
by Horja in \cite{Horj:DX,Horj:EZ}.\footnote{I thank P.~Horja for
confirming that his construction is consistent with (\ref{eq:EtoZ}).}
We refer to these references for more details.

Given the form of the monodromy (\ref{eq:EtoZ}) it is easy to
reproduce all the corresponding results of section \ref{s:two} for
this example.

%%%%%%%%%%%%%%%%%%%
\subsection{A reducible exceptional divisor} \label{ss:red}

One might have got the impression from the localization argument in
section \ref{ss:C2} that we can understand the monodromy associated to
an orbifold singularity by studying the little \CY\ $W$ living inside the
exceptional divisor $E$. Indeed this argument shows that the analysis we
did in section \ref{s:two} for a \CY\ threefold would also apply
locally to a
\CY\ fivefold which has a $\Z_{18}$ orbifold singularity given by
the action
\begin{equation}
  (z_1,z_2,z_3,z_4,z_5)\mapsto(\alpha^9z_1,\alpha^6z_2,\alpha z_3,
	\alpha z_4,\alpha z_5),
\end{equation}
where $\alpha=\exp(2\pi i/18)$. This is because the resolved
$\P^4_{\{9,6,1,1,1\}}$ is the exceptional divisor for this
five-dimensional orbifold.

While this is useful in some circumstances it does not mean that any
orbifold analysis can be reduced to a \CY\ computation in lower
dimensions. The problem is that the exceptional divisor for an
orbifold singularity may be {\em reducible\/}. Indeed one generically
expects an exceptional divisor be to reducible. In this case the
notion of the \CY\ ``inside'' the exceptional divisor makes no sense.

At least in the context of toric cases we can make some general
statements about the difference between a reducible and irreducible
exceptional divisor. The impression one might have been left with from
the above examples is that each particular exceptional divisor $E$ is
associated with its own component $\Delta_E$ of the discriminant
divisor. In this case one then associates monodromy around $\Delta_E$
with a process involving the collapse of $E$.

\def\ptsA{\mathcal{A}}
In the Batyrev \cite{Bat:m} way of describing \CY\ $n$-folds, one has
a set of points, $\ptsA$, lying in a hyperplane in $\R^{n+2}$. Vectors
from the origin to these points generate the one-dimensional edges of
a fan. By the usual algorithm in toric geometry this fan gives
the canonical line bundle over some $(n+1)$-dimensional variety $V$. $X$
is then the ``little \CY'' living inside $V$.

Let $Q$ by the convex hull of $\ptsA$.
One can show (see chapter 10 of \cite{GKZ:book}) that the irreducible
components of the discriminant are determined by
faces of $Q$ of various dimensions.
For a nontrivial component we require at
least $k+2$ points in a $k$-dimensional face.

If each face of $Q$ has at most one point in its interior then each
divisor associated to this interior point gets its 
own component of the discriminant. This was the case for the examples
studied above and was the case for all the examples studied in
\cite{GJ:McK,Tomas:McK,Mayr:McK}. It is precisely when we have a
reducible exceptional divisor that this fails.

We will consider an example of this.  For a change we will use a K3
surface rather 
than a \CY\ threefold. The results generalize easily to the
threefold case.

Consider the surface $X$ of degree 12 in $\P^3_{\{4,3,3,2\}}$. 
This has $\C^2/\Z_2$ singularities at 3 points and $\C^2/\Z_3$
singularities at 4 points. Each $\Z_2$ singularity is resolved by a
single $\P^1$ and each $\Z_3$ singularity is resolved by a sum of two
$\P^1$'s intersecting at a point. The fact that the K3 is embedded in
the given ambient space ties the blow-ups of the points of similar
type together. There are in fact
only four K\"ahler degrees of freedom --- the overall size of the
initial weighted projective space, one blow-up of all the $\Z_2$
fixed points and two blow-ups for all the $\Z_3$ fixed points.

\begin{table}
\def\m{\phantom{-}}
\[\begin{array}{|c|cccc|c|}
\hline
a_0&\m0&\m0&\m0&1&-2H\\
a_1&\m0&\m0&\m1&1&B\\
a_2&\m1&\m0&\m0&1&C\\
a_3&\m1&\m2&\m0&1&C\\
a_4&-3&-3&-2&1&-H+A\\
a_5&-2&-2&-1&1&2H-2A+B\\
a_6&-1&-1&\m0&1&A-2B\\
a_7&\m1&\m1&\m0&1&H-2C\\
\hline
\end{array}\]
\caption{The point-set $\ptsA$ for the K3 example.}   \label{tab:K3}
\end{table}

We list the coordinates of the points $\ptsA$ in
table~\ref{tab:K3}. We also show the 
divisor classes in terms of a basis $(H,A,B,C)$ associated to these
vectors. These divisor classes restrict to form a partial basis of
$H_2(X)=H^2(X)$. We then demand that the K\"ahler form of $X$ lies in the
span of these generators. This basis has been chosen so that the
resulting slice of the K\"ahler cone is the positive orthant.

We will only concern ourselves with monodromies associated to the
$\Z_3$ blow-ups. To do this we only consider variations in the $A$ and
$B$ components of the K\"ahler form. The two divisor classes of
interest are those associated to $a_5$ and $a_6$ having classes
$2H-2A+B$ and $A-2B$ respectively. These each intersect $X$ in four 
copies of $\P^1$. Each $\P^1$ from one set intersects one member
from the other set of $\P^1$'s in one point. Together these form  the
resolutions of the four $\Z_3$ fixed 
points. Note that
the set $\{a_4,a_5,a_6,a_1\}$ lies on a straight line.

Let us fix algebraic coordinates
\begin{equation}
  x=\frac{a_4a_6}{a_5^2},\quad y=\frac{a_1a_5}{a_6^2}.
\end{equation}
In terms of these, the component of the discriminant associated with
the line $\{a_4,a_5,a_6,a_1\}$ is
\begin{equation}
  \Delta_1=27x^2y^2-18xy-1+4x+4y.  \label{eq:dZ3}
\end{equation}
There are two other components of the discriminant --- the primary
component and one associated to the three $\Z_2$ fixed
points. $\Delta_1$ is the part intrinsically associated to the
$\C^2/\Z_3$ resolution.

\iffigs
\begin{figure}
  \centerline{\epsfxsize=7cm\epsfbox{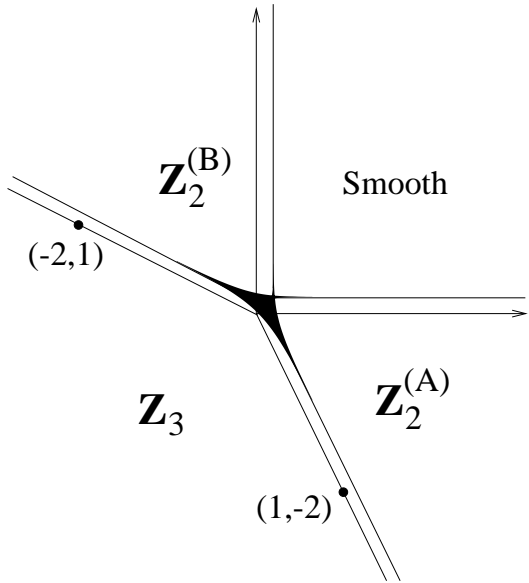}}
  \caption{The $\C^2/\Z_3$ resolution phase space.}
  \label{fig:Z3}
\end{figure}
\fi

The discriminant (\ref{eq:dZ3}) divides the $(-\log x,-\log y)$ into
four phases as shown in figure~\ref{fig:Z3}. One has the resolved phase, the
$\Z_3$ orbifold phase and two phases where one of the pair of $\P^1$'s
has been blown up to partially resolve the $\Z_3$ fixed point to
something that looks locally like a $\Z_2$ fixed point. We will call
the latter two phases $\Z_2^{(A)}$ and $\Z_2^{(B)}$ since they are
associated to blowing down using the $A$ or $B$ component of the
K\"ahler form respectively.

Note again that there is only a {\em
single\/} irreducible component, $\Delta_1$, of the divisor associated
to this picture. The same component of the discriminant is responsible
for blowing down either of the $\P^1$'s.

Now, of course, the moduli space is really four dimensional. We want
to think of figure~\ref{fig:Z3} as representing a slice of the moduli
space, where the $H$ and $C$ components of the K\"ahler form have been
taken to infinity. Equivalently, think of figure~\ref{fig:Z3} as
the toric fan of a two-dimensional subspace of the moduli space
associated to this limit.

Let $\O_{a_5}$ be the sum of the four structure sheaves of the
$\P^1$'s associated to $a_5$. Similarly $\O_{a_6}$ is supported only
over the four $\P^1$'s associated to $a_6$.
Now consider the $\P^1$ in the moduli space connecting the large
radius limit to the $\Z_2^{(A)}$ limit point where the $\P^1$'s
associated to $a_5$ are blown down. Let us denote this by $C_A$.
We know the monodromy around the large radius limit within $C_A$
multiplies the Chern characters by $e^A$. Since $\Delta_1$ hits $C_A$
transversely we expect that monodromy around the discriminant point is
given by $\cK^K_{\O_{a_5}}$ which induces
\begin{equation}
  \ch(\cF) \mapsto \ch(\cF)-\ff14\langle\O_{a_5},\cF\rangle\O_{a_5},
\end{equation}
where the factor $\ff14$ appears because $\O_{a_5}$ is associated to
{\em four\/} irreducible divisors.\footnote{It is not hard to convince
yourself that such a factor is necessary to get the monodromies to
come out correctly. It would be nice to explain this factor more
completely. Presumably this is similar to a asking for a better
understanding of (\ref{eq:EtoZ}).}

The $\Z_2^{(A)}$ limit point is associated to a $\Z_2$ orbifold point
and so we expect going twice around this point gives something
simple. One can show that applying $\cK^K_{\O_{a_5}}\cK^B_A$ twice
induces multiplication by $\exp(2H+B)$ on the Chern characters. This
should be viewed in the same way as section \ref{ss:C2}.

What about the curve $C_B$ which connects the large radius limit
to the $\Z_2^{(B)}$ limit point? In this case the discriminant
point induces monodromy
given by $\cK^K_{\O_{a_6}}$ which induces
\begin{equation}
  \ch(\cF) \mapsto \ch(\cF)-\ff14\langle\O_{a_6},\cF\rangle\O_{a_6}.
\end{equation}
One can then show that applying $\cK^K_{\O_{a_6}}\cK^B_B$ twice
induces multiplication by $\exp(A)$ on the Chern characters. 

This is all very well except that $\cK^K_{\O_{a_5}}$ and
$\cK^K_{\O_{a_6}}$ are both monodromies around the {\em same\/}
component of the discriminant. The reason they are different is that
if we fix a base point near the large radius limit then the loop
around $\Delta_1$ inside $C_A$ is not homotopic to the loop around
$\Delta_1$ inside $C_B$. This can happen because $\Delta_1$ itself is
not smooth. It has a cusp at the point $(x,y)=(\ff13,\ff13)$. If we
draw an $S^3$ around this cusp then we obtain a trefoil knot in the
intersection. It is well-known that, for a fixed basepoint, loops
around different parts of this knot are not homotopic to each
other. This allows for the difference between $\cK^K_{\O_{a_5}}$ and
$\cK^K_{\O_{a_6}}$. Note that this is the very same cusp as the one
studied by Argyres and Douglas \cite{DA:SU3}.

Finally let us see how to compute the effect of monodromy going three
times around the $\Z_3$ limit point. Let $C_{A2}$ be the curve
connecting the $\Z_2^{(A)}$ limit point to the $\Z_3$ limit point. The
monodromy within $C_{A2}$ around the $\Z_2^{(A)}$ limit point is
identical to the monodromy within $C_{A}$ around the $\Z_2^{(A)}$
limit point for the reasons given in section \ref{ss:C3}. The
$\Delta_1$ component of the discriminant intersects $C_{A2}$
transversely. Since this is associated with collapsing the divisor
associated to $a_6$ we will assume that monodromy around this
discriminant point is given by $\cK^K_{\O_{a_6}}$. {\em We therefore claim
that monodromy with $C_A$ around the $\Z_3$ limit point is given by 
$\cK^K_{\O_{a_6}}\cK^K_{\O_{a_5}}\cK^B_A$.}

We may check that this cubes to something nice. Indeed the effect on
the Chern characters implies that
\begin{equation}
(\cK^K_{\O_{a_6}}\cK^K_{\O_{a_5}}\cK^B_A)^3 = (\cK^B_{H})^4.
\end{equation}
One can show that this is consistent with the global geometry of the
moduli space and so our monodromies all have the expected properties.

%%%%%%%%%%%%%%%%%%%%%%%%%%%%%%%%%%%%%%%%%%%%%%%%%%%%%%%%%%%%%%%%%%%

\section{Discussion of the 0-Brane} \label{s:D0}

We have given various rules for how to compute the effect of
monodromy on the Chern character of a D-brane. In this last section we
will discuss the consequences for a 0-brane.

The 0-brane is of particular interest as it is the basic object used
in the construction of Bondal and Orlov \cite{BO:DCeq} to build the
target space from the derived category. The fact that the 0-brane can
transform into something else under monodromy is one reason why the
Bondal and Orlov construction is ambiguous for a \CY\ space.

Note again that the following results could
have been guessed using the Picard--Fuchs differential equations. In
that language the 0-brane often appears as a constant solution to the
differential equations \cite{AGM:sd}. The derived category provides a
much simpler picture however.

For a \CY\ threefold $X$, let $P$ be an object in $\DC(X)$ which
corresponds to a 0-brane. This immediately implies that $\ch(P)=p$,
where $p\in H^6(X)$ is Poincar\'e dual to a point. Under monodromy
about $\Delta_0$ we have
\begin{equation}
\begin{split}
  p &\mapsto p + \int_X p\wedge\td(X)\\
	&= p + 1.
\end{split}
\end{equation}
That is, the 0-brane always picks up a 6-brane charge upon an orbit around
$\Delta_0$.

Now consider the other monodromies in this paper. They all involve
taking the structure sheaf $\O_E$ of some collapsing cycle $E$ and
computing $\langle\O_E,p\rangle$. If $E$ is of dimension less than 6
then this inner product is always zero. Thus the 0-brane undergoes no
monodromy about these kinds of components in the discriminant.

We therefore make the following
\begin{conjecture}
The 0-brane undergoes monodromy if and only if we circle the primary
component, $\Delta_0$, of the discriminant.
\end{conjecture}

If we begin in a large radius smooth \CY\ phase, which other phases
may we visit without crossing a wall in the phase diagram which
contains $\Delta_0$? In other words, over what area of the phase
diagram can we fix a choice of 0-brane without worrying about monodromy?
The answer consists of the so-called ``geometric phases'' or
``partially enlarged K\"ahler moduli space'' of \cite{AGM:II}.

The phases correspond to triangulations of Batyrev's reflexive
polytope \cite{Bat:m,AGM:II}. The statement that a phase is geometric
corresponds to every 
simplex in the triangulation having the unique point in the interior of
the polytope as a vertex.

Comparing to the example in section \ref{s:two} for example, the
geometric phases consist of the \CY\ phase and the orbifold phase
where we have a three complex dimensional picture of the target
space. Indeed, these phases are separated only by $\Delta_1$ around
which the 0-brane has no monodromy.

The geometric phases consist of those reached from the \CY\ phase only
by blowing down subspaces. If one reduces the overall dimension of
the target space then one must cross a wall containing
$\Delta_0$. There are also exotic ``exoflop'' transitions
\cite{AGM:II} where part of the target space remains three-dimensional
but a lower-dimensional part grows out of the side of the target
space. These exoflops also involve crossing a $\Delta_0$ boundary and
are not considered geometric.

Proving the statement about these phases is an application of the
combinatorics discussed in chapter 11 of \cite{GKZ:book}. There is an
object $\eta_T$ which is a function on the cones of the secondary
fan. If $\eta_T$ changes as you pass to a neighbouring cone then the
wall contained $\Delta_0$. It is easy to show that $\eta_T$ changes as
you pass from a geometric phase to a non-geometric phase. This result
is essentially contained in corollary 4.5 of chapter 11 of
\cite{GKZ:book}. It is a more tedious exercise to show that passing between
geometric phases of threefolds keeps $\eta_T$ constant.

This result appears to jibe nicely with the Bondal and Orlov
construction. We may consistently tie the derived category to a target
space interpretation so long as we confine ourselves to geometric
phases. Once we leave these phases then the 0-brane undergoes
monodromy and we acquire an ambiguity in the way we construct the
target space.

Finally we should note that there can still be an ambiguity in what
exactly is called a 0-brane in the geometric phases. If there are two
or more smooth phases related by flops then each phase has its own
0-branes. The exact way these branes are related to each other was
given by Bridgeland \cite{Brig:flop}.

%%%%%%%%%%%%%%%%%%%%%%%%%%%%%%%%%%%%%%%%%%%%%%%%%%%%%%%%%%%%%%%%%%%

\section*{Acknowledgments}

It is a pleasure to thank M.~Douglas, A.~Lawrence, D.~Morrison, R.~Plesser,
E.~Sharpe, M.~Vybornov and M.~Stern for useful conversations. I am
particularly grateful to P.~Horja for giving me an advanced copy of
\cite{Horj:EZ} while this paper was being prepared, and discussing the
results with me. 
The author is supported in part by NSF grant DMS-0074072 and a research
fellowship from the Alfred P.~Sloan Foundation. The author is also
grateful to the ITP, Santa Barbara for hospitality while this paper
was being completed with support from NSF grant PHY99-07949.

%\bibliographystyle{my-phys}
%\bibliography{string}

\end{document}